\documentclass[%
 reprint,
superscriptaddress,
 amsmath,amssymb,
 aps,
prl,
]{revtex4-2}
\usepackage{mathtools}
\usepackage{graphicx}
\usepackage{dcolumn}
\usepackage{bm}
\usepackage{hyperref}
\usepackage{xr-hyper}
\usepackage{cleveref}
\usepackage{color}
\usepackage[dvipsnames]{xcolor}
\usepackage[justification=justified]{caption}
\usepackage{subcaption}
\usepackage{ragged2e}
\usepackage[inkscapelatex=false]{svg}

\DeclareCaptionJustification{justified}{\justifying}
\begin{document}

\preprint{APS/123-QED}

\title{Exploiting Negative Capacitance for Unconventional Coulomb Engineering}
 \author{Aravindh Shankar*}
\email{abhavani@purdue.edu}
\author{Pramey Upadhyaya*}
\email{prameyup@purdue.edu}
\author{Supriyo Datta*}
\email{datta@purdue.edu}
\affiliation{*All authors contributed equally\\Elmore Family School of Electrical and
Computer Engineering, Purdue University, West Lafayette,
Indiana 47907, USA}
\date{\today}

\begin{abstract}
The many-body ground state of a two-dimensional electron system can be tuned by Coulomb engineering through control of the dielectric environment. However, in conventional dielectrics the static permittivity is restricted to positive values, limiting the accessible interaction regimes. Here we argue that the negative capacitance demonstrated in appropriately engineered structures can open new vistas for Coulomb engineering. The associated negative permittivity could  transform the natural repulsive interaction of electrons into an attractive one, raising the intriguing possibility of nontrivial ground states, including superconductivity. Using models of two-dimensional electron systems with linear and parabolic dispersion relations coupled to environments with negative capacitance, we estimate the strength and sign of the engineered Coulomb interaction and outline parameter regimes that could stabilize correlated electronic phases.
\end{abstract}

\maketitle

\textit{Introduction.}|Coulomb interactions between electrons govern a variety of many-body phenomena, from stabilizing interacting phases to shaping the electrical and optical responses of matter. Consequently, several approaches have been developed to highlight the effects driven by Coulomb interactions, bringing them to the forefront of condensed matter physics. For example, one prominent approach~\cite{bistritzer2011moire} focuses on revealing Coulomb interactions by suppressing competing kinetic energies in materials engineered to host flat energy bands. This technique has successfully stabilized interacting phases ranging from superconductivity~\cite{cao2018,lu2019superconductors,balents2020superconductivity,oh2021} to the fractional quantum anomalous Hall state~\cite{sharpe2019qahe,serlin2020qahe,li2021qahe} in two-dimensional materials~\cite{kennes2021moirereview}. A complementary strategy|known as \textit{Coulomb engineering}|enables the direct tuning of Coulomb interactions by designing heterostructures with tailored electromagnetic environments~\cite{raja2017,song2018}. Although Coulomb engineering has successfully altered the electrical and optical responses of materials~\cite{jang2008graph,hwang2012fvr,yu2013qcap,kajino2019coulomb,whelan2020fvr}, it remains challenging to tune Coulomb interactions over a wide range and to design interaction-driven phases of matter within this framework. This limitation arises because previously explored electromagnetic environments are largely restricted to positive static permittivities. However, on fundamental grounds, environments with \textit{local} static negative permittivity are permissible \cite{dolgov1981}, raising the question of whether present-day Coulomb-engineered structures can be extended beyond this constraint.

Here we answer this question in the affirmative by proposing and studying novel Coulomb-engineered structures motivated by recent demonstrations of ferroelectric negative capacitance (NC)~\cite{zubko2016,yadav2019,das2021local}. Specifically, we propose to modify the structures in those works to embed two-dimensional electron systems (2DES) such that the normally repulsive interaction between electrons is rendered attractive, thereby unveiling  a new regime of unconventional Coulomb engineering. On the one hand, this extended tuning of Coulomb interactions can change the transport properties and optical response in the normal state of 2DES, potentially exceeding limits of conventional Coulomb engineering~\cite{jang2008graph,hwang2012fvr,yu2013qcap,raja2017,song2018,kajino2019coulomb,whelan2020fvr}. On the other hand, entirely new phases of matter may emerge~\cite{van2023cemott}, especially once the customarily repulsive electron–electron interactions are rendered attractive. 

While predicting the ground state is beyond the scope of the present work, a state of particular interest with attractive electron interactions is superconductivity, which requires the formation of paired electronic states~\cite{bcs1957}. Motivated by this, we estimate the pairing strength in a model platform where attractive interactions are mediated by negative capacitance arising from the motion of ferroelectric domain walls~\cite{luk2018prb}. Within this model, the induced attraction is retarded by the ferroelectric domain wall response, whose characteristic frequencies typically lie well below electronic energy scales, leading to a separation of timescales analogous to the phonon-mediated pairing. We find that the resulting dimensionless pairing parameter $\lambda - \mu^{*}$, adapted from electron–phonon theory~\cite{tolmachev1961,morelanderson,colemanbook}, can reach values $\gtrsim 0.1$ in realistic experimental regimes. Crucially, by balancing positive and negative capacitances, this parameter is highly tunable within our proposal, suggesting that NC-enabled Coulomb engineering could provide a new route towards designing superconductivity and other ordered electronic phases. More broadly, there has recently been growing interest in cavity-based materials engineering~\cite{riolo2025polariton,lu2025cavity}, where the collective response of specially designed cavities is leveraged to modify the ground state of solid-state materials. Our results indicate that the collective response of ferroelectrics stabilized in the negative-capacitance regime could offer a useful resource for this paradigm.

\begin{figure}
    \centering
    \includegraphics[width=0.47\textwidth]{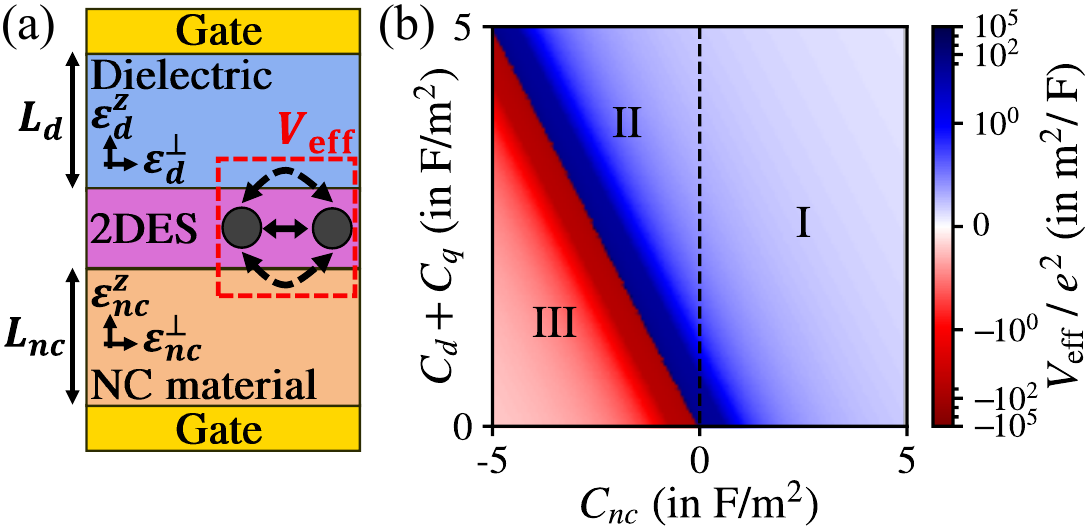}
    \caption{Central idea.~(a) System schematic: a two-dimensional electron system (2DES) surrounded by dielectric (conventional) and negative-capacitance (NC) media, with respective thicknesses $L_{d}$ and $L_{nc}$. When the NC material is a ferroelectric, this structure is referred to as Metal-Ferroelectric-2DES-Insulator-Metal (MF2IM) configuration.
    (b) Long-wavelength part of engineered 2DES Coulomb interaction $V_{\text{eff}}$ as per Eq.~(\ref{Eq3}), shown as a function of geometric capacitances $C_{nc}$ and $C_{d}$, and quantum capacitance $C_{q}$. The black dashed line indicates the limit of conventional Coulomb engineering ($C_{nc}>0$, region I). Region II is new but expected to be unstable in the MF2IM configuration. Notably, $V_{\text{eff}}$ can become $\textit{negative}$ in region III, where thermodynamic stability is satisfied as per Eq.~(\ref{Eq4}).}
    \label{Fig1}
\end{figure}

\textit{Central Idea.}|To illustrate the main concept, we begin with the idealized structure shown schematically in Fig.~\ref{Fig1}(a) (connections to real material platforms are made in later sections). A two-dimensional electron system (2DES) is sandwiched between a gated conventional dielectric (DE) of thickness $L_{d}$ and a gated NC material of thickness $L_{nc}$. We model the DE and NC regions with diagonal permittivity tensors $\varepsilon^{nc(d)} = \text{diag}(\varepsilon^{nc(d)}_{\perp}, \varepsilon^{nc(d)}_{\perp}, \varepsilon^{nc(d)}_{z})$, where $z$ denotes the stacking direction and the superscripts $nc$ and $d$ represent NC and DE regions, respectively. In the 2DES plane, the static Coulomb interaction $V_{\text{eff}}$ as a function of in-plane wavenumber $q$ is given by:
\begin{equation}
V_{\mathrm{eff}}(q)=\frac{v_q}{\varepsilon_{\mathrm{eff}}(q)},\qquad
v_q\equiv \frac{e^2}{2\varepsilon_0 q}.
\label{Eq1}
\end{equation}
We model the effective dielectric function $\varepsilon_{\mathrm{eff}}(q)$ using linear response polarization functions as: $\varepsilon_{\mathrm{eff}}(q) = 1 - v_{q}(\Pi_{nc}(q) + \Pi_{d}(q) + \Pi_{el}(q))$. The environment dielectric function $\varepsilon_{\mathrm{env}}(q) \equiv 1 - v_{q}(\Pi_{nc}(q) + \Pi_{d}(q))$ can be derived from Poisson's equation (see Appendix A) and reflects the `gate screening' effect:
\begin{equation}
2\varepsilon_{\text{env}}(q) = \frac{\sqrt{\vphantom{\varepsilon^{d}}\varepsilon^{nc}_{\perp}\varepsilon^{nc}_{z}}}{\tanh\Big(q\sqrt{\vphantom{\varepsilon^{d}}\varepsilon^{nc}_{\perp}/\varepsilon^{nc}_{z}}L_{nc}\Big)} + \frac{\sqrt{\varepsilon^{d}_{\perp}\varepsilon^{d}_{z}}}{\tanh\Big(q\sqrt{\varepsilon^{d}_{\perp}/\varepsilon^{d}_{z}}L_{d}\Big)}.
    \label{Eq2}
\end{equation}
Introducing dimensionless anisotropy factors $\eta_{nc} = \sqrt{\varepsilon^{nc}_{\perp}/\varepsilon^{nc}_{z}}$ and $\eta_{d} = \sqrt{\varepsilon^{d}_{\perp}/\varepsilon^{d}_{z}}$, we illustrate the central idea of NC-enabled unconventional Coulomb engineering by first considering the limit where $q\eta_{nc}L_{nc} \ll 1$ and $q\eta_{d}L_{d} \ll 1$. In this regime, the wavelengths associated with low-energy scattering of electrons in the 2DES are large compared to the gate distances and  characteristic length scales of the polarizable DE and NC media (elaborated in later sections). For such large wavelengths ($q\to 0$), Eqs.~(\ref{Eq1})--(\ref{Eq2}) reduce to a simple form:
\begin{equation}
    V_{\text{eff}} \cong \frac{e^{2}}{C_{nc} + C_{d} + C_{q}},
    \label{Eq3}
\end{equation}
where $C_{nc}\equiv\varepsilon_{0}\varepsilon^{nc}_{z}/{L_{nc}}$, $C_{d}\equiv \varepsilon_{0}\varepsilon^{d}_{z}/{L_{d}}$ are geometric capacitances per unit area, and $C_{q}\equiv-e^{2}\Pi_{el}(q\to 0)$ is the quantum capacitance of the 2DES. The expression for $C_{q}$ reflects the compressibility sum rule~\cite{giuliani2008quantum}. For conventional dielectrics, $C_{d}>0$, whereas $C_{nc}$ can be negative for materials operating in the NC regime. Throughout this work we focus on the case of $C_{q}>0$, which may be violated in certain low-density regimes of a 2DES due to exchange and correlation effects~\cite{eisenstein1992negative}. 

Before discussing the scope of this result, we first present a stability condition which constrains the values $C_{nc}$ can assume in the proposed configuration. The NC state of a material in isolation is thermodynamically unstable, but as argued in Ref.~\cite{salahuddindatta2008}, it can be locally stabilized in a constituent layer provided the composite structure is stable against charge fluctuations. We refer to the structure in Fig.~\ref{Fig1}(a) as Metal-Ferroelectric-2DES-Insulator-Metal (MF2IM) configuration (specifying NC material as ferroelectric to connect with notation used in literature~\cite{park2019modeling,luk2022ferroelectric}). The necessary and sufficient condition for stability of NC in the MF2IM configuration is (assuming $C_{nc} < 0$, and $C_{d}, C_{q} > 0$, see Appendix B for details):
\begin{equation}
    C_{nc} + C_{d} + C_{q} < 0.
    \label{Eq4}
\end{equation}
Setting $C_{q} = 0$ recovers the familiar stability condition for the widely studied Metal-Ferroelectric-Insulator-Metal (MFIM) structure~\cite{salahuddindatta2008,park2019modeling,iniguez2019ferroelectric}. In the opposite limit of $C_{q}\to\infty$, the MF2IM structure decouples into two separate capacitors (assuming the 2DES is connected to a charge reservoir), and NC cannot be stabilized in this case. 

Fig.~\ref{Fig1}(b) illustrates the first main result of this work by treating the capacitances as free variables. In conventional Coulomb engineering, where all capacitances are positive, the tunability of $V_{\text{eff}}$ is limited to region I. The inclusion of $C_{nc}<0$ opens additional regimes, labeled regions II and III. According to Eq.~(\ref{Eq4}), region II is unstable against charge fluctuations. If the system is tuned into this region|for example, by varying $C_{q}$|the charge distribution will reconfigure to restore stability. This behavior may provide a new diagnostic tool for probing NC materials. In contrast, region III remains stable while reversing the \textit{sign} of $V_{\mathrm{eff}}$, converting the intrinsically repulsive electron–electron interaction into an effective attraction. Moreover, as per Eq.~(\ref{Eq3}), the magnitude of this attractive $V_{\mathrm{eff}}$ can be engineered to reach large values through careful matching of the constituent capacitances. In the remainder of this work, we focus on region III, emphasizing model experimental platforms capable of realizing this regime.

\begin{figure}
    \centering
    \includegraphics[width=0.475\textwidth]{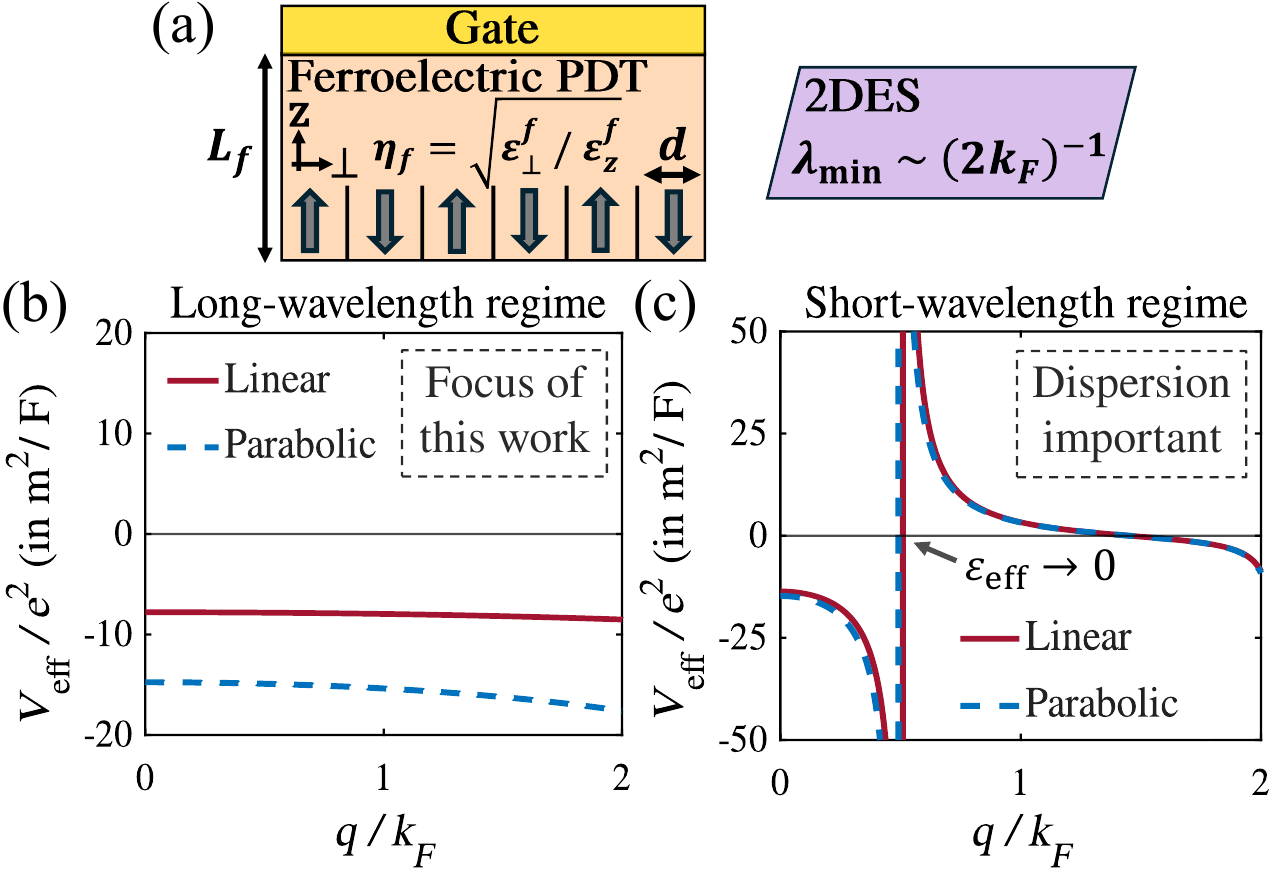}
    \caption{Critical length scales.~(a) The wavelength of low-energy electron scattering in the 2DES, $(2k_{F})^{-1}$, relative to $2d$ (twice the domain width) in ferroelectric periodic domain texture (PDT) governs whether dispersion $\varepsilon^{f}_{z}(q)$ is important in extrinsic NC case, and relative to $|\eta_{f}|L_{f}$ (scaled ferroelectric thickness) determines whether the linear approximation of gate screening effect is valid. In the intrinsic NC case, $2d$ should be replaced by an appropriate lattice-related length scale. (b), (c) Illustration of effective Coulomb interaction energy $V_{\text{eff}}$ as a function of wavenumber $q$ as per Eqs.~(\ref{Eq1})--(\ref{Eq2}), for linear and parabolic 2DES with typical parameter values $v_{F}=1\times 10^{6}~\text{m}/\text{s}$ and $m^{*} = 0.1m_{e}$ respectively. Here, 2DES carrier density $n$ was varied to access the long-wavelength (low $n = 10^{15}/\text{m}^{2}$) and short-wavelength (high $n = 10^{17}/\text{m}^{2}$) regimes for fixed $L_{f} = 4~\text{nm}$.}
    \label{Fig2}
\end{figure}

\begin{figure}
    \centering
    \includegraphics[width=0.47\textwidth]{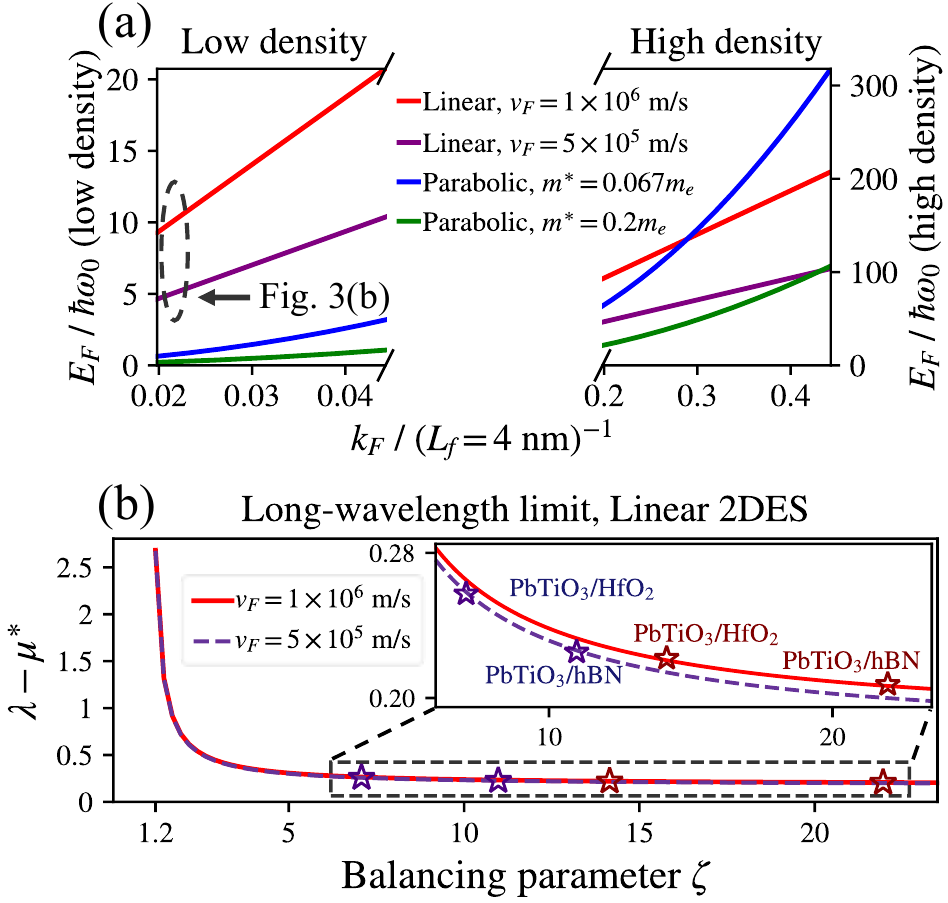}
       \caption{Critical energy scales and pairing strength.~(a) Fermi energy $E_{F}$ of the 2DES, expressed in units of PDT single-mode energy $\hbar\omega_{0}$ for ferroelectric $\text{PbTiO}_{3}$, plotted as a function of Fermi wavenumber $k_{F}$ normalized to the ferroelectric thickness $L_{f}$. The long-wavelength regime described in Fig.~\ref{Fig2} is accessed in the limit of low density, shown as circled region. (b) Pairing strength parameter $\lambda - \mu^{*}$ calculated for a linear 2DES in the long-wavelength approximation as a function of balancing parameter $\zeta$ defined in Eq.~(\ref{Eq11}), shown for two typical values of Fermi velocity $v_{F}$. Here, carrier density was fixed at $n = 1\times 10^{15} / \text{m}^{2}$. Inset highlights where typical material systems (NC/DE) fall, according to the models used in this work.}
    \label{Fig3}
\end{figure}

\textit{Model Platform.}| Here, we focus on ferroelectrics as the specific choice of NC material. Within the growing literature on negative capacitance, one distinguishes transient NC~\cite{khan2015transient,khan2017transient,hoffmann2018transient,hoffmann2019unveiling,kim2019transient,qiao2021both}, which occurs during polarization switching and reflects nonequilibrium order-parameter relaxation, from stabilized NC~\cite{appleby2014stable,gao2014stable,zubko2016,hoffmann2018stabilization,yadav2019,das2021local,qiao2021both,qiao2024stable}, corresponding to a quasi-static regime with negative permittivity. NC systems are further classified by material class|perovskites~\cite{yadav2019}, oxides~\cite{hoffmann2022intrinsic}, van der Waals materials~\cite{wang2019vdw}|and by ferroic order, including ferroelectrics ~\cite{iniguez2019ferroelectric} and antiferroelectrics~\cite{qiao2021both,hoffmann2022antiferroelectric}. Here we focus on stabilized NC, distinguishing extrinsic NC~\cite{zubko2016,yadav2019,das2021local}, associated with multidomain configurations and domain-wall motion~\cite{junquera2023topological}, from intrinsic NC~\cite{hoffmann2022intrinsic}, which appears independent of domain structure. While the general framework presented in Fig.~\ref{Fig1} is agnostic to the underlying mechanism, our material-specific calculations (Figs.~\ref{Fig2} and \ref{Fig3}) employ a domain-based model of extrinsic NC, which we now detail for stabilized NC in perovskite ferroelectrics, exemplified by $\text{PbTiO}_{3}$.
This choice is motivated by substantial evidence from experiments~\cite{zubko2016,yadav2019,das2021local} and corresponding theoretical explanation~\cite{luk2018prb,luk2019harnessing} providing a framework for modeling near-equilibrium NC. 

In ferroelectric $\text{PbTiO}_{3}$ ($\varepsilon^{nc}_{z(\perp)}\to\varepsilon^{f}_{z(\perp)}$, $L_{nc}\to L_{f}$), the polarization response is highly anisotropic, with $\varepsilon^{f}_{z} < 0$ and $\varepsilon^{f}_{\perp} > 0$. Starting from a periodic domain texture (PDT) as the equilibrium polarization configuration, the static negative permittivity $\varepsilon^{f}_{z}$ is explained as an overscreening effect due to the role of depolarization field in this system~\cite{luk2018prb,luk2019harnessing}. Assuming small deviations from equilibrium, the PDT dynamics can be described by an oscillator model for domain wall displacements, where $P$ is the electric polarization and $E_{\text{tot}}$ is the total electric field (external + depolarization):
\begin{equation}
    \ddot{P}(t) + (\omega_{0}^{2} - \Omega^{2})P(t) = \Omega^{2}\varepsilon_{0}\varepsilon^{f}_{z,hf}E_{\text{tot}}(t).
    \label{Eq5}
\end{equation}
The restoring force originates from Coulomb energy associated with excess surface charges when the PDT is displaced from equilibrium. Here, $\omega_{0}$ is the characteristic oscillation frequency of the PDT system and $\Omega$ represents the strength of coupling to external electric field. We set the damping~\cite{luk2018prb} to zero, as we focus on the static response in this work. In a more detailed model, the frequency parameters should exhibit dispersion $\omega_{0}(q), \Omega(q)$ with additional indices for excitation type (longitudinal vs transverse) and branch (acoustic vs optic). The expressions which we use to calculate $\omega_{0}, \Omega$ in this work (see Appendix C) account for only the longitudinal optic branch at zero wavenumber, based on the expectation that this mode contributes dominantly to the NC effect in the long-wavelength limit (defined below). We refer to this as the `single-mode' approximation.

The static linear response dielectric function corresponding to Eq.~(\ref{Eq5}) is:
\begin{equation}
    \varepsilon^{f}_{z} \equiv \varepsilon^{f}_{z}(q=0,\omega=0) = \frac{\varepsilon^{f}_{z,\mathrm{hf}}\,\omega_{0}^{2}}{\omega_{0}^{2} - \Omega^{2}}.
    \label{Eq6}
\end{equation}
Here $\varepsilon^{f}_{z,hf}>0$ denotes the high-frequency dielectric constant (see Appendix C for more details). Since $\omega_{0}<\Omega$ for ferroelectric $\text{PbTiO}_{3}$ in this model, the expression in Eq.~(\ref{Eq6}) yields a  \textit{negative} $\varepsilon^{f}_{z}$.

For the conventional dielectric part of environment (see Fig.~\ref{Fig1}(a)), we adopt values for hexagonal boron nitride
(hBN) as a representative `low-k' dielectric. Specifically, we set $\varepsilon^{d}_{z} = 3.4$, $\varepsilon^{d}_{\perp} = 6.86$ based on values reported in literature~\cite{ohbahbn,laturiahbn,Pierrethbn,kimhbn}. In Fig.~\ref{Fig3}(b), we choose hafnium oxide ($\text{HfO}_{2}$) as a representative `high-k' dielectric in the long-wavelength regime with $\varepsilon^{d}_{z} = 25$. The dielectric thickness is fixed at $L_{d} = 4~\text{nm}$ in all cases.

The bare 2DES (in vacuum, $\varepsilon_{\text{env}} = 1$) is modeled as a single-valley non-interacting electron gas, with either linear or parabolic low-energy dispersion. The two cases are characterized respectively by a Fermi velocity $v_{F}$, or an effective mass $m^{*}$. Spin and valley degeneracies are set to $g_{s} = 2$, $g_{v} = 1$ respectively. For the electronic screening, we use the Thomas-Fermi (TF) model and set $v_{q}\Pi_{el} = -k_{\mathrm{TF}}/q$, where the TF screening wavenumber $k_{\text{TF}}$~\cite{dassarma2011review} is given by:
\begin{equation}
k_{\mathrm{TF}}=
\frac{e^2}{\pi\hbar}\times
\begin{cases}
k_F/v_F, & \text{Linear},\\
m^*/\hbar, & \text{Parabolic}.
\end{cases}
\label{Eq7}
\end{equation}
Neglecting exchange and correlation effects, the compressibility sum rule is satisfied with $C_{q}=2\varepsilon_{0}k_{\mathrm{TF}}$. We set the `background' dielectric constant of 2DES to unity for simplicity.

\textit{Results.}|We begin by analyzing the Fourier-transformed Coulomb interaction in the model platform introduced above. To this end, we first note that the quantity $\eta_{nc} \equiv \sqrt{\varepsilon^{nc}_{\perp}/\varepsilon^{nc}_{z}}$ is \textit{imaginary} for an anisotropic NC material like $\text{PbTiO}_{3}$ which has $\varepsilon^{f}_{z}<0~,~\varepsilon^{f}_{\perp}>0$. Consequently the $tanh$ function in Eq.~(\ref{Eq2}) becomes a $tan$ function ($\tanh(ix) = i\tan(x)$):
\begin{equation}
    \varepsilon_{\mathrm{env}}(q) = \frac{1}{2}\bigg[\frac{|\eta_{f}|\varepsilon^{f}_{z}}{\tan\big(q\hspace{0.03cm}|\eta_{f}|L_{f}\big)} +
\frac{\eta_{d}\varepsilon^{d}_{z}}{\tanh\big(q\hspace{0.05cm}\eta_{d}L_{d}\big)}\bigg],
\label{Eq8}
\end{equation}
where $|\eta_{f}|$ denotes the complex magnitude, and we have assumed that the dielectric part is conventional ($\varepsilon^{d}_{z}, \varepsilon^{d}_{\perp} > 0$). This form reveals that in the case of anisotropic NC the calculated $V_{\text{eff}}$ is periodic, and it is natural to identify the corresponding critical length scale $|\eta_{f}|L_{f}$. For a system with isotropic NC ($\varepsilon^{f}_{z}<0$, $\varepsilon^{f}_{\perp}<0$) $\eta_{f}$ is real-valued, the corresponding gate screening function remains $tanh$, and $V_{\text{eff}}(q)$ is not periodic.

In addition to $|\eta_{f}|L_{f}$, two important length scales in the composite structure are $2k_{F}$ (low-energy scattering shell in the 2DES) and a characteristic length scale corresponding to the NC phenomenon, taken to be the domain width $d$ of PDT in the extrinsic case. In Fig.~\ref{Fig2}, we illustrate the definition of two distinct regimes based on the relative values of these length scales. In the long-wavelength limit|$2k_{F}d\ll 1$ and $2k_{F}|\eta_{f}|L_{f}\ll 1$|two simplifications emerge: wavelengths of electron scattering in 2DES are too large to resolve microscopic details on the scale of domain width $d$ (so that a model with constant $\varepsilon^{nc}_{z/\perp}(q)$ suffices), and gate screening reduces to its linear limit since the gates see an effectively uniform charge distribution in the 2DES. In this regime, the single-mode approximation discussed in previous section is a reasonable starting point, and the simplified description of Eq.~(\ref{Eq3}) is applicable [see Fig.~\ref{Fig2}(b)]. In the short-wavelength regime|$2k_{F}d\gg 1$ and $2k_{F}|\eta_{f}|L_{f}\gg 1$|it is important to model the dispersion $\varepsilon^{f}_{z}(q)$ accurately (single-mode approximation may not be reliable) and gate screening modifies the $q$-space structure of $V_{\text{eff}}(q)$ significantly [see Fig.~\ref{Fig2}(c)]. 

In frequency space, the negative permittivity associated with the PDT response [Eq.~(\ref{Eq5})] persists below the characteristic PDT oscillation frequency $\omega_0$~\cite{luk2018prb}. Consequently, the Coulomb interaction mediated by the NC ferroelectric remains attractive for $\omega<\omega_0$. When the Fermi energy satisfies $E_F \gg \hbar \omega_0$, this separation of energy scales allows the retarded attractive interaction to partially overcome the instantaneous Coulomb repulsion, analogous to phonon-mediated pairing. Motivated by this observation, we next estimate the pairing strength within our model platform.

Fig.~\ref{Fig3}(a) shows the ratio of 2DES Fermi energy $E_{F}$ to PDT oscillation energy $\hbar\omega_{0}$ (for $4~\text{nm}$ thick $\text{PbTiO}_{3}$) as a function of carrier density. At low density, this condition begins to be satisfied ($E_{F}\sim 10\hbar\omega_{0}$) only for linear 2DES with high $v_{F}$. At high density, both linear and parabolic 2DES comfortably satisfy this condition ($E_{F}\sim 100\hbar\omega_{0}$) for typical $v_{F}, m^{*}$ values. The theory developed in this work is most applicable in the low-energy, long-wavelength regime. For this reason, we focus in the remainder of this section on the case of linear 2DES at low density, which we model in the long-wavelength approximation using Eq.~(\ref{Eq3}). From the perspective of real systems, the high density regime and case of parabolic 2DES are just as likely to exhibit interesting physics, but careful treatment of $V_{\text{eff}}(q,\omega)$ in the short-wavelength, high-energy regime is required and beyond the scope of this work.

To quantify the strength of low-energy electron pairing in the 2DES, we use parameters $\lambda$ and $\mu^{*}$, adapted from the theory of electron-phonon interaction and superconductivity~\cite{colemanbook}. To explain their definition in our context, we begin by decomposing the self-consistently screened interaction $V_{\text{eff}}$ in the following manner, based on the expected slowness of NC-mediated screening relative to direct screening by electrons ($\hbar\omega_{0}\ll E_{F}$):
\begin{align}
    \begin{aligned}
        V_{\text{eff}}(q) &= v_{q} + v_{q}\Pi_{\text{el}}(q)V_{\text{eff}}(q) + v_{q}\Pi_{\text{env}}(q)V_{\text{eff}}(q)\\
        &= \hspace{0.8cm}\underbrace{V_{\text{el}}(q)}_{\text{fast}} \hspace{0.8cm} + \underbrace{V_{\text{el}}(q)\Pi_{\text{env}}(q)V_{\text{eff}}(q)}_{\text{slow}},
    \end{aligned}
    \label{Eq9}
\end{align}
where $\Pi_{\text{env}}(q) = \Pi_{nc}(q) + \Pi_{d}(q)$ and $V_{\text{el}}(q) = v_{q}/(1 - v_{q}\Pi_{\text{el}}(q)$). We call the `slow' term in Eq.~(\ref{Eq9}) $\Tilde{V}_{\text{env}}$ and calculate the following dimensionless constants~\cite{colemanbook}:
\begin{equation}
        \lambda = -N_{0}\langle \Tilde{V}_{\text{env}}\rangle\hspace{0.045cm};\hspace{0.045cm}\mu = N_{0}\langle V_{el}\rangle\hspace{0.045cm};\hspace{0.045cm}\mu^{*} = \frac{\mu}{1 + \mu \ln(\frac{E_{F}}{\hbar\omega_{0}})},
    \label{Eq10}
\end{equation}
where angle brackets denote averaging over Fermi surface (FS), $q(\theta)$ is appropriately defined for the 2D system, and $N_{0}$ is the density of states at $E_{F}$ per spin per unit area. The Coulomb pseudopotential $\mu^{*}$ \cite{morelanderson} represents a renormalization of direct Coulomb repulsion ($\mu$) due to the large bandwidth for electronic screening ($\sim E_{F}$) compared to environment-mediated screening ($\sim\hbar\omega_{0}$). Carrier pairing at accessible temperatures is expected when $\lambda - \mu^{*} > 0$ and sufficiently large ($\gtrsim 0.1$).

In the long-wavelength limit, the problem can be fully specified by fixing values for the 2DES $n$ and $v_{F}$, and introducing a balancing parameter $\zeta$ defined as follows:
\begin{equation}
    C_{nc} + C_{d} = -\zeta C_{q}.
    \label{Eq11}
\end{equation}
To satisfy the thermodynamic stability condition in Eq.~(\ref{Eq4}) we assume $\zeta > 1$. Fig.~\ref{Fig3}(b) shows calculated values of $\lambda - \mu^{*}$ for NC $\text{PbTiO}_{3}$ and linear 2DES ($n = 1\times 10^{15}~/\text{m}^{2}$) in the long-wavelength limit. We emphasize two points from this figure. First, estimates for the model platform introduced above yield $\lambda-\mu^{*}\gtrsim0.1$ (inset). Although these estimates rely on simplifying assumptions, discussed further in the next section, they suggest that pairing strengths of experimentally relevant magnitude may be achievable. For comparison, in monolayer graphene the value of $\lambda$ arising from other pairing mechanisms such as the electron-phonon interaction is many orders of magnitude smaller~\cite{einenkel2011graphene}. Second, in contrast to phonon-mediated pairing where the coupling strength is typically fixed by material properties, balancing the capacitances of NC, DE and 2DES offers a new route to \textit{engineer} the value of $\lambda - \mu^{*}$. Particularly, as the balancing parameter approaches $\zeta = 1$ from above, pairing can be pushed into strong coupling regimes\footnote{We note that the assumptions of linear response and the existence of a small parameter $\lambda \hbar \omega_0/E_F$, which underlie the present pairing estimates, may break down as $\zeta \rightarrow 1$. This could ultimately limit the achievable pairing strength.}.

\textit{Discussion.}|
In summary, we have introduced a new method for Coulomb engineering using negative-capacitance materials. We show that incorporating NC media enables the effective electron–electron interaction in two-dimensional systems to be tuned and even reversed in sign  while maintaining thermodynamic stability (Fig.~\ref{Fig1}). We additionally present a stability condition for NC in the MF2IM configuration [Eq.~(\ref{Eq4})], categorization of Coulomb engineering into regimes based on critical length scales(Fig.~\ref{Fig2}), and estimates of interaction strengths for realistic material platforms (Fig.~\ref{Fig3}). We now outline the limitations of our analysis and discuss directions for future work.

A more quantitative description of $V_{\mathrm{eff}}(\vec{q},\omega)$ requires extending the present treatment beyond the long-wavelength and low-frequency limits considered here. For the ferroelectric NC medium, this includes: (i) incorporating dispersion of the optic branch $\omega_{+}(q)$ and coupling $\Omega(q)$ to extend the present results for $\varepsilon_z^f(q=0)$ to finite $q$ (see Appendix C); (ii) including contributions from all excitation branches in the PDT system~\cite{sidorkin2006domain}; (iii) accounting for anisotropy introduced by domain structures, which may lead to angular dependence of $\epsilon_z^f(\vec{q})$ and influence pairing symmetry; (iv) extending the description of NC beyond the Kittel model~\cite{yang2021condensation}, particularly given experimentally observed vortex-like structures~\cite{yadav2019}; and (v) incorporating the temperature dependence of the NC phase and its dielectric response~\cite{pavlenko2021temperature}.

For the 2DES, further refinements include incorporating finite-thickness corrections through $q$-dependent form factors~\cite{AndoFowlerStern1982} and extending $\Pi_{\rm el}$ beyond the static, zero-temperature limit by using finite-temperature dynamical screening approaches~\cite{giuliani2008quantum}. This extension incorporates the plasmon, which is important even in the low-frequency limit due to its gapless nature in two dimensions. Hybridization~\cite{veld2023screening} of this mode with excitations of the NC medium can modify $V_{\mathrm{eff}}(\vec{q},\omega)$ and conclusions about the resulting ground states.

Finally, the estimates of pairing strength using Eq.~(\ref{Eq10}) can be improved by analyzing the full momentum- and frequency-dependent interaction $V(\vec{q},\omega)$ within a Green's function framework~\cite{{Eliashberg1960,einenkel2011graphene}}. When pair-breaking processes limit superconductivity, then the transition temperature scales as $T_c \sim \exp(-1/(\lambda - \mu^{*}))$, so that small errors in estimating the quantity $\lambda - \mu^{*}$ can cause large errors in estimating $T_c$. This makes it important to design experiments capable of measuring small values of $\lambda - \mu^{*}$ directly. Moreover, in two dimensions $T_{c}$ may instead be limited by phase fluctuations, highlighting the need to estimate the superfluid stiffness~\cite{randeria2021limits} in future work.

\textit{Acknowledgements.}|SD would like to thank Sayeef Salahuddin for alerting him to transient NC and for helpful discussions regarding negative permittivity in ferroelectrics. He would also like to thank Bhaskaran Muralidharan, Kerem Camsari and Zubin Jacob for their feedback on an earlier version of the manuscript. AS and PU gratefully acknowledge helpful discussions with Allan H. MacDonald on anisotropic gate screening. This work was supported by the Purdue Research Foundation and the Blue Sky Research Program through the Purdue College of Engineering.

\appendix
\section{APPENDIX}
\section{A. Anisotropic gate screening effect with Negative Capacitance} \label{AppA}
We consider a two-dimensional electron system (2DES) encapsulated on one side by a dielectric medium with homogeneous positive static permittivity tensor, 
$\varepsilon^{d} = \mathrm{diag}(\varepsilon_{\perp}^{d}, \varepsilon_{\perp}^{d}, \varepsilon_{z}^{d})$, 
and on the other side by a negative-capacitance (NC) medium described by 
$\varepsilon^{nc} = \mathrm{diag}(\varepsilon_{\perp}^{nc}>0, \varepsilon_{\perp}^{nc}>0, \varepsilon_{z}^{nc}<0)$, 
where $z$ denotes the stacking direction.

From Poisson’s equation, the electrostatic potential $\varphi$ in the two encapsulating regions takes the form:
\begin{equation}
\begin{gathered}
    \varphi(q,z>0) = \alpha^{(d)}e^{\eta_{d}qz} + \beta^{(d)}e^{-\eta_{d}qz},\\
    \varphi(q,z<0) = \alpha^{(nc)}e^{\eta_{nc}qz} + \beta^{(nc)}e^{-\eta_{nc}qz},
\end{gathered}
\label{Eq1supp}
\end{equation}
with anisotropy factors $\eta_{d}$ = $\sqrt{\varepsilon^{d}_{\perp}/\varepsilon^{d}_{z}}$ and $\eta_{nc}$ = $\sqrt{\varepsilon^{nc}_{\perp}/\varepsilon^{nc}_{z}}$. Metal gates are placed at distances $L_{d}, L_{nc}$ above and below the $z = 0$ plane where the 2DES is located.

Imposing the boundary condition $\varphi(q,L_{d/nc})$ = 0 at the gates and enforcing continuity of $\varphi$ across the 2DES plane, we obtain:
\begin{align}
\begin{aligned}
    \varphi(q) \equiv \varphi(q,z = 0) &= \alpha^{(nc)}(1 - e^{-2\eta_{nc}qL_{nc}}) \\&= \beta^{(d)}(1 - e^{-2\eta_{d}qL_{d}}).
\end{aligned}
\label{Eq2supp}
\end{align}
The potential due to a test charge in the 2DES is determined by applying the condition for electric displacement field, $\hat{n}.(\vec{D}_{1} - \vec{D}_{2}) = \sigma$, across the 2DES with charge density $\sigma$ = e$\delta(\vec{\rho}=0)$. Here, $\vec{\rho}$ is the two-dimensional position vector. Taking Fourier transform of both sides gives:
\begin{align}
\begin{aligned}
    \alpha^{(nc)}&\varepsilon_{z}^{nc}\partial_{z}(e^{\eta_{nc}qz} - e^{-2\eta_{nc}qL_{nc}}e^{-\eta_{nc}qz})\big\vert_{z=0} \\- &\beta^{(d)}\varepsilon_{z}^{d}\partial_{z}(e^{-\eta_{d}qz} - e^{-2\eta_{d}qL_{d}}e^{\eta_{d}qz})\big\vert_{z=0} = e / \varepsilon_{0}.
\end{aligned}
    \label{Eq3supp}
\end{align}

Solving for $\alpha^{(nc)}$ and substituting in Eq.~(\ref{Eq2supp}), the potential in 2DES is:
\begin{equation}
    \varphi(q) = \frac{e}{\varepsilon_{0}}\times
        \dfrac{1}{\varepsilon_{z}^{d}(q)\dfrac{\eta_{d}q}{\tanh(\eta_{d}qL_{d})} + \varepsilon_{z}^{nc}\dfrac{\eta_{nc}q}{\tanh(\eta_{nc}qL_{nc})}}.
        \label{Eq4supp}
\end{equation}
The corresponding environment contribution to the interaction energy, 
$V_{\text{env}}(q) = e\varphi(q)$, defines the effective background permittivity via
\begin{equation}
    V_{\text{env}}(q) = \frac{e^{2}}{2\varepsilon_{0}\varepsilon_{\text{env}}q}.
    \label{Eq5supp}
\end{equation}
Finally, we note that for anisotropic NC media $\eta_{nc}$ is purely imaginary, but the static quantities $\varphi(q)$, $V_{\text{env}}(q)$ (and $V_{\text{eff}}(q)$ in main text) remain strictly real.
\section{B. Stability of Negative Capacitance in MF2IM configuration}
\label{AppB}
We consider a system where the dielectric is held at voltage $V_{d}$, the NC medium at $V_{nc}$, and the 2DES at $V_{q}$. For small charge fluctuations, the free energy of the system is
\begin{equation}
    \mathcal{F} = \frac{Q_{d}^{2}}{2C_{d}} + \frac{Q_{nc}^{2}}{2C_{nc}} + \frac{Q_{q}^{2}}{2C_{q}} - Q_{d}V_{d} - Q_{nc}V_{nc} - Q_{q}V_{q},
    \label{Eq6supp}
\end{equation}
subject to the constraint: $Q_{d} + Q_{nc} + Q_{q} = 0$. Using this constraint to eliminate one variable, we obtain a reduced free energy $\tilde{\mathcal{F}}$, whose Hessian matrix is
\begin{align}
    \begin{aligned}
        H &= \begin{bmatrix}
        \partial_{Q_{d}}^{2}\tilde{\mathcal{F}} & \partial_{Q_{d}}\partial_{Q_{nc}}\tilde{\mathcal{F}}\\ \partial_{Q_{nc}}\partial_{Q_{d}}\tilde{\mathcal{F}} & \partial_{Q_{nc}}^{2}\tilde{\mathcal{F}}
    \end{bmatrix} \\ &= \begin{bmatrix}
        \dfrac{1}{C_{d}} + \dfrac{1}{C_{q}} & \dfrac{1}{C_{q}}\\ \dfrac{1}{C_{q}} & \dfrac{1}{C_{nc}} + \dfrac{1}{C_{q}}
    \end{bmatrix}.
    \end{aligned}
    \label{Eq7supp}
\end{align}
The necessary and sufficient condition for $H$ to be positive definite (i.e., for both eigenvalues to be strictly positive) is
\begin{equation}
    \frac{1}{C_{d}C_{nc}} + \frac{1}{C_{nc}C_{q}} + \frac{1}{C_{q}C_{d}} > 0.
    \label{Eq8supp}
\end{equation}
Assuming $C_{nc} < 0$ and $C_{d}, C_{q} > 0$, this condition simplifies to the stability condition quoted in the main text as Eq.~(4).

An important consideration neglected in this work is the modification of $C_{q}$ (and $\Pi_{el}$ in main text) by `direct' electron-electron interactions (the screening channel constituted by identical electrons within the 2DES). The strength of this interaction relative to kinetic energy is often expressed in terms of the dimensionless parameter $r_{s}$. For a 2DES with spin and valley
degeneracies $g_s$ and $g_v$ and background dielectric constant $\kappa$, we have the following expressions for different cases of dispersion relation~\cite{dassarma2011review}:
\begin{equation}
r_s =
\begin{cases}
\displaystyle \frac{e^{2}}{4\pi\varepsilon_{0}\kappa\hbar v_{F}}\,
\frac{\sqrt{g_s g_v}}{2}, & \text{Linear}, \\[6pt]
\displaystyle \frac{m^{*}e^{2}}{8\pi\varepsilon_{0}\kappa\hbar^{2}}\,
\frac{g_s g_v}{\sqrt{\pi n}}, & \text{Parabolic}.
\end{cases}
\label{Eq9supp}
\end{equation}

For a 2DES at large enough $r_{s}$, the modified $C_{q}$ could itself be negative due to exchange-correlation corrections~\cite{eisenstein1992negative,giuliani2008quantum}. In that case,  Eqs.~(\ref{Eq6supp}) and (\ref{Eq7supp}) remain valid, but Eq.~(\ref{Eq8supp}) becomes only a necessary (not sufficient) condition, and  Eq.~(4) of the main text must be revised. A more detailed analysis indicates that, within the model used in this section, an MF2IM structure with both $C_{nc}<0$ and $C_{q}<0$ would actually be unstable, in the sense that no simple necessary and sufficient stability condition can be formulated. It is nevertheless conceivable that an appropriately designed stabilizing circuit could enable access to this regime, should it prove to be experimentally relevant.

\section{C. Domain-provided Negative Capacitance model} \label{AppC}
In a ferroelectric with periodic domain texture (PDT), the `restoring force' characterizing the response of domain walls to an external electric field can be calculated within the Kittel model~\cite{kittel1946} of alternating `hard' domains. The equilibrium domain width in this model~\cite{catalan2012domain} is:
\begin{equation}
    d = \sqrt{3.53\sqrt{\frac{\varepsilon^{f}_{\perp}}{\varepsilon^{f}_{z,hf}}}\xi\delta L_{f}},
    \label{Eq10supp}
\end{equation}
where $\xi = 2\times(1 + \varepsilon^{d}_{z}/\sqrt{\varepsilon^{f}_{\perp}\varepsilon^{f}_{z,hf}})$ is a parameter capturing electrostatic boundary conditions, $\delta \approx 1~\text{nm}$ is the domain wall thickness~\cite{luk2018prb}, and $L_{f}$ is the ferroelectric thickness.

The PDT system has a `stiffness' because there is an energy cost from long-range Coulomb forces when this system is displaced from equilibrium. The oscillation frequency corresponding to this stiffness is calculated in the Kittel model~\cite{sidorkin2006domain} using Fourier analysis:
\begin{align}
\begin{aligned}
    \omega_{\pm}^{2}&(q) = \frac{4P_{s}^{2}}{\pi\varepsilon_{0}\sqrt{\varepsilon^{f}_{\perp}\varepsilon^{f}_{z}}ML_{f}}\times \\ &\Bigg[\sum_{n=1}^{\infty}\textrm{ln}\bigg(1 + \frac{\varepsilon^{f}_{\perp}}{\varepsilon^{f}_{z}}\frac{L_{f}^{2}}{d^{2}}\frac{1}{(2n-1)^{2}}\bigg)\big(1 \pm \cos((2n-1)qd)\big) \\ &- \sum_{n=1}^{\infty}\textrm{ln}\bigg(1 + \frac{\varepsilon^{f}_{\perp}}{\varepsilon^{f}_{z}}\frac{L_{f}^{2}}{d^{2}}\frac{1}{(2n)^{2}}\bigg)(1 - \cos(2nqd))\Bigg],
\end{aligned}
    \label{Eq11supp}
\end{align}
where the two branches ($\pm$) of excitation for this system are analogous to the acoustic and optic branches in crystals with a two-atom basis. In this work we restrict attention to the $+$ (`optic') branch in the long-wavelength limit, $q\to 0$. We calculate $\omega_{0}$ in the main text as $\omega_{+}(q=0)$ and assume electric dipole coupling of the form $-P_{z}\times E_{z}$~\cite{sidorkin2006domain, luk2018prb}:
\begin{align}
\begin{aligned}
    \omega_{0}^{2} = \frac{8P_{s}^{2}}{\pi\varepsilon_{0}\sqrt{\varepsilon^{f}_{\perp}\varepsilon^{f}_{z,\text{hf}}}ML_{f}}&\mathrm{ln}\Bigg(\mathrm{cosh}\Bigg(\sqrt{\frac{\varepsilon^{f}_{\perp}}{\varepsilon^{f}_{z,\text{hf}}}}\frac{\pi}{2}\frac{L_{f}}{d}\Bigg)\Bigg)\\
    \Omega^{2} = 
        &\dfrac{4P_{s}^{2}}{\varepsilon_{0}\varepsilon^{f}_{z,\text{hf}}Md}.
\end{aligned}
\label{Eq12supp}
\end{align}

Parameters for ferroelectric $\text{PbTiO}_{3}$ are chosen as follows~\cite{luk2018prb} : $P_{s} = 0.65~\text{Cm}^{-2}$ is the spontaneous polarization, and $M$ is the domain wall mass per unit area estimated using interpolation formula $M = 1.3\sqrt{L_{f}[\text{nm}]}\times 10^{-9}~\text{kgm}^{-2}$~\cite{luk2018prb,zhangponomareva}. $\varepsilon^{f}_{z,hf} = 100$ is the high-frequency dielectric constant capturing the background polarizability of $\text{PbTiO}_{3}$ arising from other polarization mechanisms in the material, and $\varepsilon^{f}_{\perp}=30$.

Within this model, the calculated negative permittivity $\varepsilon^{f}_{z}$ has a value of about $-60$ for $L_{f} = 4~\text{nm}$.


\bibliographystyle{apsrev4-2}
\bibliography{apssamp.bib}

@article{bistritzer2011moire,
  title={Moir{\'e} bands in twisted double-layer graphene},
  author={Bistritzer, Rafi and MacDonald, Allan H},
  journal={Proceedings of the National Academy of Sciences},
  volume={108},
  number={30},
  pages={12233--12237},
  year={2011},
  publisher={National Acad Sciences}
}

@article{cao2018,
  title={Unconventional superconductivity in magic-angle graphene superlattices},
  author={Cao, Yuan and Fatemi, Valla and Fang, Shiang and Watanabe, Kenji and Taniguchi, Takashi and Kaxiras, Efthimios and Jarillo-Herrero, Pablo},
  journal={Nature},
  volume={556},
  number={7699},
  pages={43--50},
  year={2018},
  publisher={Nature Publishing Group}
}

@article{lu2019superconductors,
  title={Superconductors, orbital magnets and correlated states in magic-angle bilayer graphene},
  author={Lu, Xiaobo and Stepanov, Petr and Yang, Wei and Xie, Ming and Aamir, Mohammed Ali and Das, Ipsita and Urgell, Carles and Watanabe, Kenji and Taniguchi, Takashi and Zhang, Guangyu and others},
  journal={Nature},
  volume={574},
  number={7780},
  pages={653--657},
  year={2019},
  publisher={Nature Publishing Group UK London}
}

@article{balents2020superconductivity,
  title={Superconductivity and strong correlations in moir{\'e} flat bands},
  author={Balents, Leon and Dean, Cory R and Efetov, Dmitri K and Young, Andrea F},
  journal={Nature Physics},
  volume={16},
  number={7},
  pages={725--733},
  year={2020},
  publisher={Nature Publishing Group UK London}
}

@article{oh2021,
  title={Evidence for unconventional superconductivity in twisted bilayer graphene},
  author={Oh, Myungchul and Nuckolls, Kevin P and Wong, Dillon and Lee, Ryan L and Liu, Xiaomeng and Watanabe, Kenji and Taniguchi, Takashi and Yazdani, Ali},
  journal={Nature},
  volume={600},
  number={7888},
  pages={240--245},
  year={2021},
  publisher={Nature Publishing Group UK London}
}

@article{sharpe2019qahe,
  title={Emergent ferromagnetism near three-quarters filling in twisted bilayer graphene},
  author={Sharpe, Aaron L and Fox, Eli J and Barnard, Arthur W and Finney, Joe and Watanabe, Kenji and Taniguchi, Takashi and Kastner, MA and Goldhaber-Gordon, David},
  journal={Science},
  volume={365},
  number={6453},
  pages={605--608},
  year={2019},
  publisher={American Association for the Advancement of Science}
}

@article{serlin2020qahe,
  title={Intrinsic quantized anomalous Hall effect in a moir{\'e} heterostructure},
  author={Serlin, Marec and Tschirhart, CL and Polshyn, Hryhoriy and Zhang, Yuxuan and Zhu, Jiacheng and Watanabe, Kenji and Taniguchi, Takashi and Balents, L and Young, AF},
  journal={Science},
  volume={367},
  number={6480},
  pages={900--903},
  year={2020},
  publisher={American Association for the Advancement of Science}
}

@article{li2021qahe,
  title={Quantum anomalous Hall effect from intertwined moir{\'e} bands},
  author={Li, Tingxin and Jiang, Shengwei and Shen, Bowen and Zhang, Yang and Li, Lizhong and Tao, Zui and Devakul, Trithep and Watanabe, Kenji and Taniguchi, Takashi and Fu, Liang and others},
  journal={Nature},
  volume={600},
  number={7890},
  pages={641--646},
  year={2021},
  publisher={Nature Publishing Group UK London}
}

@article{kennes2021moirereview,
  title={Moir{\'e} heterostructures as a condensed-matter quantum simulator},
  author={Kennes, Dante M and Claassen, Martin and Xian, Lede and Georges, Antoine and Millis, Andrew J and Hone, James and Dean, Cory R and Basov, DN and Pasupathy, Abhay N and Rubio, Angel},
  journal={Nature Physics},
  volume={17},
  number={2},
  pages={155--163},
  year={2021},
  publisher={Nature Publishing Group UK London}
}

@article{raja2017,
  title={Coulomb engineering of the bandgap and excitons in two-dimensional materials},
  author={Raja, Archana and Chaves, Andrey and Yu, Jaeeun and Arefe, Ghidewon and Hill, Heather M and Rigosi, Albert F and Berkelbach, Timothy C and Nagler, Philipp and Sch{\"u}ller, Christian and Korn, Tobias and others},
  journal={Nature communications},
  volume={8},
  number={1},
  pages={15251},
  year={2017},
  publisher={Nature Publishing Group UK London}
}

@article{song2018,
  title={Electron quantum metamaterials in van der Waals heterostructures},
  author={Song, Justin CW and Gabor, Nathaniel M},
  journal={Nature nanotechnology},
  volume={13},
  number={11},
  pages={986--993},
  year={2018},
  publisher={Nature Publishing Group UK London}
}

@article{jang2008graph,
  title={Tuning the effective fine structure constant in graphene: Opposing effects of dielectric screening on short-and long-range potential scattering},
  author={Jang, C and Adam, Shaffique and Chen, J-H and Williams, Ellen D and Das Sarma, S and Fuhrer, MS},
  journal={Physical review letters},
  volume={101},
  number={14},
  pages={146805},
  year={2008},
  publisher={APS}
}

@article{hwang2012fvr,
  title={Fermi velocity engineering in graphene by substrate modification},
  author={Hwang, Choongyu and Siegel, David A and Mo, Sung-Kwan and Regan, William and Ismach, Ariel and Zhang, Yuegang and Zettl, Alex and Lanzara, Alessandra},
  journal={Scientific reports},
  volume={2},
  number={1},
  pages={590},
  year={2012},
  publisher={Nature Publishing Group UK London}
}

@article{whelan2020fvr,
  title={Fermi velocity renormalization in graphene probed by terahertz time-domain spectroscopy},
  author={Whelan, Patrick R and Shen, Qian and Zhou, Binbin and Serrano, Ismael G and Kamalakar, M Venkata and Mackenzie, David MA and Ji, Jie and Huang, Deping and Shi, Haofei and Luo, Da and others},
  journal={2D Materials},
  volume={7},
  number={3},
  pages={035009},
  year={2020},
  publisher={IOP Publishing}
}

@article{yu2013qcap,
  title={Interaction phenomena in graphene seen through quantum capacitance},
  author={Yu, GL and Jalil, R and Belle, Branson and Mayorov, Alexander S and Blake, Peter and Schedin, Frederick and Morozov, Sergey V and Ponomarenko, Leonid A and Chiappini, F and Wiedmann, S and others},
  journal={Proceedings of the National Academy of Sciences},
  volume={110},
  number={9},
  pages={3282--3286},
  year={2013},
  publisher={National Acad Sciences}
}

@article{kajino2019coulomb,
  title={Modification of optical properties in monolayer WS2 on dielectric substrates by coulomb engineering},
  author={Kajino, Yuto and Oto, Kenichi and Yamada, Yasuhiro},
  journal={The Journal of Physical Chemistry C},
  volume={123},
  number={22},
  pages={14097--14102},
  year={2019},
  publisher={ACS Publications}
}

@article{van2023cemott,
  title={Coulomb engineering of two-dimensional Mott materials},
  author={van Loon, Erik GCP and Sch{\"u}ler, Malte and Springer, Daniel and Sangiovanni, Giorgio and Tomczak, Jan M and Wehling, Tim O},
  journal={npj 2D Materials and Applications},
  volume={7},
  number={1},
  pages={47},
  year={2023},
  publisher={Nature Publishing Group UK London}
}

@article{dolgov1981,
  title = {On an admissible sign of the static dielectric function of matter},
  author = {Dolgov, O. V. and Kirzhnits, D. A. and Maksimov, E. G.},
  journal = {Rev. Mod. Phys.},
  volume = {53},
  issue = {1},
  pages = {81--93},
  numpages = {0},
  year = {1981},
  month = {Jan},
  publisher = {American Physical Society}
}

@article{riolo2025polariton,
  title={Polariton-induced superconductivity in two-dimensional metals},
  author={Riolo, Riccardo and Koppens, Frank HL and Jarillo-Herrero, Pablo and Mazza, Giacomo and MacDonald, Allan H and Polini, Marco},
  journal={arXiv preprint arXiv:2511.00608},
  year={2025}
}

@article{lu2025cavity,
  title={Cavity engineering of solid-state materials without external driving},
  author={Lu, I-Te and Shin, Dongbin and Kamper Svendsen, Mark and Latini, Simone and H{\"u}bener, Hannes and Ruggenthaler, Michael and Rubio, Angel},
  journal={Advances in Optics and Photonics},
  volume={17},
  number={2},
  pages={441--525},
  year={2025},
  publisher={Optica Publishing Group}
}

@article{zubko2016,
  title={Negative capacitance in multidomain ferroelectric superlattices},
  author={Zubko, Pavlo and Wojde{\l}, Jacek C and Hadjimichael, Marios and Fernandez-Pena, St{\'e}phanie and Sen{\'e}, Ana{\"\i}s and Luk’yanchuk, Igor and Triscone, Jean-Marc and {\'I}{\~n}iguez, Jorge},
  journal={Nature},
  volume={534},
  number={7608},
  pages={524--528},
  year={2016},
  publisher={Nature Publishing Group UK London}
}

@article{yadav2019,
  title={Spatially resolved steady-state negative capacitance},
  author={Yadav, Ajay K and Nguyen, Kayla X and Hong, Zijian and Garc{\'\i}a-Fern{\'a}ndez, Pablo and Aguado-Puente, Pablo and Nelson, Christopher T and Das, Sujit and Prasad, Bhagwati and Kwon, Daewoong and Cheema, Suraj and others},
  journal={Nature},
  volume={565},
  number={7740},
  pages={468--471},
  year={2019},
  publisher={Nature Publishing Group UK London}
}

@article{das2021local,
  title={Local negative permittivity and topological phase transition in polar skyrmions},
  author={Das, Sujit and Hong, Zijian and Stoica, VA and Gon{\c{c}}alves, MAP and Shao, Yu-Tsun and Parsonnet, Eric and Marksz, Eric J and Saremi, Sahar and McCarter, MR and Reynoso, A and others},
  journal={Nature materials},
  volume={20},
  number={2},
  pages={194--201},
  year={2021},
  publisher={Nature Publishing Group UK London}
}

@article{iniguez2019ferroelectric,
  title={Ferroelectric negative capacitance},
  author={{\'I}{\~n}iguez, Jorge and Zubko, Pavlo and Luk’yanchuk, Igor and Cano, Andr{\'e}s},
  journal={Nature Reviews Materials},
  volume={4},
  number={4},
  pages={243--256},
  year={2019},
  publisher={Nature Publishing Group UK London}
}

@article{bcs1957,
  title={Theory of superconductivity},
  author={Bardeen, John and Cooper, Leon N and Schrieffer, John Robert},
  journal={Physical review},
  volume={108},
  number={5},
  pages={1175},
  year={1957},
  publisher={APS}
}

@inproceedings{tolmachev1961,
  title={Logarithmic criterion for superconductivity},
  author={Tolmachev, Vladimir Veniaminovich},
  booktitle={Doklady Akademii Nauk},
  volume={140},
  pages={563--566},
  year={1961},
  organization={Russian Academy of Sciences}
}

@article{morelanderson,
  title={Calculation of the superconducting state parameters with retarded electron-phonon interaction},
  author={Morel, P and Anderson, PW},
  journal={Physical Review},
  volume={125},
  number={4},
  pages={1263},
  year={1962},
  publisher={APS}
}

@book{colemanbook,
  title={Introduction to many-body physics},
  author={Coleman, Piers},
  year={2015},
  publisher={Cambridge University Press}
}

@article{salahuddindatta2008,
  title={Use of negative capacitance to provide voltage amplification for low power nanoscale devices},
  author={Salahuddin, Sayeef and Datta, Supriyo},
  journal={Nano letters},
  volume={8},
  number={2},
  pages={405--410},
  year={2008},
  publisher={ACS Publications}
}

@article{park2019modeling,
  title={Modeling of negative capacitance in ferroelectric thin films},
  author={Park, Hyeon Woo and Roh, Jangho and Lee, Yong Bin and Hwang, Cheol Seong},
  journal={Advanced Materials},
  volume={31},
  number={32},
  pages={1805266},
  year={2019},
  publisher={Wiley Online Library}
}

@article{luk2022ferroelectric,
  title={The ferroelectric field-effect transistor with negative capacitance},
  author={Luk’yanchuk, I and Razumnaya, A and Sen{\'e}, Ana{\"\i}s and Tikhonov, Y and Vinokur, VM},
  journal={npj Computational Materials},
  volume={8},
  number={1},
  pages={52},
  year={2022},
  publisher={Nature Publishing Group UK London}
}

@article{luk2019harnessing,
  title={Harnessing ferroelectric domains for negative capacitance},
  author={Luk’yanchuk, Igor and Tikhonov, Y and Sene, Anais and Razumnaya, A and Vinokur, VM},
  journal={Communications Physics},
  volume={2},
  number={1},
  pages={22},
  year={2019},
  publisher={Nature Publishing Group UK London}
}

@article{khan2015transient,
  title={Negative capacitance in a ferroelectric capacitor},
  author={Khan, Asif Islam and Chatterjee, Korok and Wang, Brian and Drapcho, Steven and You, Long and Serrao, Claudy and Bakaul, Saidur Rahman and Ramesh, Ramamoorthy and Salahuddin, Sayeef},
  journal={Nature materials},
  volume={14},
  number={2},
  pages={182--186},
  year={2015},
  publisher={Nature Publishing Group UK London}
}

@article{khan2017transient,
  title={Differential voltage amplification from ferroelectric negative capacitance},
  author={Khan, Asif I and Hoffmann, Michael and Chatterjee, Korok and Lu, Zhongyuan and Xu, Ruijuan and Serrao, Claudy and Smith, Samuel and Martin, Lane W and Hu, Chenming and Ramesh, Ramamoorthy and others},
  journal={Applied Physics Letters},
  volume={111},
  number={25},
  year={2017},
  publisher={AIP Publishing}
}

@article{hoffmann2018transient,
  title={Ferroelectric negative capacitance domain dynamics},
  author={Hoffmann, Michael and Khan, Asif Islam and Serrao, Claudy and Lu, Zhongyuan and Salahuddin, Sayeef and Pe{\v{s}}i{\'c}, Milan and Slesazeck, Stefan and Schroeder, Uwe and Mikolajick, Thomas},
  journal={Journal of Applied Physics},
  volume={123},
  number={18},
  year={2018},
  publisher={AIP Publishing}
}

@article{hoffmann2019unveiling,
  title={Unveiling the double-well energy landscape in a ferroelectric layer},
  author={Hoffmann, Michael and Fengler, Franz PG and Herzig, Melanie and Mittmann, Terence and Max, Benjamin and Schroeder, Uwe and Negrea, Raluca and Lucian, Pintilie and Slesazeck, Stefan and Mikolajick, Thomas},
  journal={Nature},
  volume={565},
  number={7740},
  pages={464--467},
  year={2019},
  publisher={Nature Publishing Group UK London}
}

@article{kim2019transient,
  title={Transient negative capacitance effect in atomic-layer-deposited Al2O3/Hf0. 3Zr0. 7O2 bilayer thin film},
  author={Kim, Keum Do and Kim, Yu Jin and Park, Min Hyuk and Park, Hyeon Woo and Kwon, Young Jae and Lee, Yong Bin and Kim, Han Joon and Moon, Taehwan and Lee, Young Hwan and Hyun, Seung Dam and others},
  journal={Advanced Functional Materials},
  volume={29},
  number={17},
  pages={1808228},
  year={2019},
  publisher={Wiley Online Library}
}

@article{appleby2014stable,
  title={Experimental observation of negative capacitance in ferroelectrics at room temperature},
  author={Appleby, Daniel JR and Ponon, Nikhil K and Kwa, Kelvin SK and Zou, Bin and Petrov, Peter K and Wang, Tianle and Alford, Neil M and O’Neill, Anthony},
  journal={Nano letters},
  volume={14},
  number={7},
  pages={3864--3868},
  year={2014},
  publisher={ACS Publications}
}

@article{gao2014stable,
  title={Room-temperature negative capacitance in a ferroelectric--dielectric superlattice heterostructure},
  author={Gao, Weiwei and Khan, Asif and Marti, Xavi and Nelson, Chris and Serrao, Claudy and Ravichandran, Jayakanth and Ramesh, Ramamoorthy and Salahuddin, Sayeef},
  journal={Nano letters},
  volume={14},
  number={10},
  pages={5814--5819},
  year={2014},
  publisher={ACS Publications}
}

@article{hoffmann2018stabilization,
  title={On the stabilization of ferroelectric negative capacitance in nanoscale devices},
  author={Hoffmann, Michael and Pe{\v{s}}i{\'c}, Milan and Slesazeck, Stefan and Schroeder, Uwe and Mikolajick, Thomas},
  journal={Nanoscale},
  volume={10},
  number={23},
  pages={10891--10899},
  year={2018},
  publisher={Royal Society of Chemistry}
}

@article{qiao2021both,
  title={Observation of negative capacitance in antiferroelectric PbZrO3 Films},
  author={Qiao, Leilei and Song, Cheng and Sun, Yiming and Fayaz, Muhammad Umer and Lu, Tianqi and Yin, Siqi and Chen, Chong and Xu, Huiping and Ren, Tian-Ling and Pan, Feng},
  journal={Nature Communications},
  volume={12},
  number={1},
  pages={4215},
  year={2021},
  publisher={Nature Publishing Group UK London}
}

@article{wang2019vdw,
  title={Van der Waals negative capacitance transistors},
  author={Wang, Xiaowei and Yu, Peng and Lei, Zhendong and Zhu, Chao and Cao, Xun and Liu, Fucai and You, Lu and Zeng, Qingsheng and Deng, Ya and Zhu, Chao and others},
  journal={Nature communications},
  volume={10},
  number={1},
  pages={3037},
  year={2019},
  publisher={Nature Publishing Group UK London}
}

@article{hoffmann2022antiferroelectric,
  title={Antiferroelectric negative capacitance from a structural phase transition in zirconia},
  author={Hoffmann, Michael and Wang, Zheng and Tasneem, Nujhat and Zubair, Ahmad and Ravindran, Prasanna Venkatesan and Tian, Mengkun and Gaskell, Anthony Arthur and Triyoso, Dina and Consiglio, Steven and Tapily, Kandabara and others},
  journal={Nature communications},
  volume={13},
  number={1},
  pages={1228},
  year={2022},
  publisher={Nature Publishing Group UK London}
}

@article{qiao2024stable,
  title={Observation of stabilized negative capacitance effect in hafnium-based ferroic films},
  author={Qiao, Leilei and Zhao, Ruiting and Song, Cheng and Zhou, Yongjian and Wang, Qian and Ren, Tian-Ling and Pan, Feng},
  journal={Materials Futures},
  volume={3},
  number={1},
  pages={011001},
  year={2024},
  publisher={IOP Publishing}
}

@article{junquera2023topological,
  title={Topological phases in polar oxide nanostructures},
  author={Junquera, Javier and Nahas, Yousra and Prokhorenko, Sergei and Bellaiche, Laurent and {\'I}{\~n}iguez, Jorge and Schlom, Darrell G and Chen, Long-Qing and Salahuddin, Sayeef and Muller, David A and Martin, Lane W and others},
  journal={Reviews of Modern Physics},
  volume={95},
  number={2},
  pages={025001},
  year={2023},
  publisher={APS}
}

@article{kittel1946,
  title={Theory of the structure of ferromagnetic domains in films and small particles},
  author={Kittel, Charles},
  journal={Physical Review},
  volume={70},
  number={11-12},
  pages={965},
  year={1946},
  publisher={APS}
}

@article{hoffmann2022intrinsic,
  title={Intrinsic Nature of Negative Capacitance in Multidomain Hf0. 5Zr0. 5O2-Based Ferroelectric/Dielectric Heterostructures},
  author={Hoffmann, Michael and Gui, Mengcheng and Slesazeck, Stefan and Fontanini, Riccardo and Segatto, Mattia and Esseni, David and Mikolajick, Thomas},
  journal={Advanced Functional Materials},
  volume={32},
  number={2},
  pages={2108494},
  year={2022},
  publisher={Wiley Online Library}
}

@book{sidorkin2006domain,
  title={Domain structure in ferroelectrics and related materials},
  author={Sidorkin, Aleksandr Stepanovich},
  year={2006},
  publisher={Cambridge Int Science Publishing}
}

@article{luk2018prb,
  title={Electrodynamics of ferroelectric films with negative capacitance},
  author={Luk'Yanchuk, I and Sene, Anais and Vinokur, VM},
  journal={Physical Review B},
  volume={98},
  number={2},
  pages={024107},
  year={2018},
  publisher={APS}
}

@article{zhangponomareva,
  title={Nanodynamics of ferroelectric ultrathin films},
  author={Zhang, Qingteng and Herchig, R and Ponomareva, I},
  journal={Physical Review Letters},
  volume={107},
  number={17},
  pages={177601},
  year={2011},
  publisher={APS}
}

@article{ohbahbn,
  title={First-principles study on structural, dielectric, and dynamical properties for three BN polytypes},
  author={Ohba, Nobuko and Miwa, Kazutoshi and Nagasako, Naoyuki and Fukumoto, Atsuo},
  journal={Physical Review B},
  volume={63},
  number={11},
  pages={115207},
  year={2001},
  publisher={APS}
}

@article{Pierrethbn, year = {2022},
month = {jun},
publisher = {IOP Publishing},
volume = {9},
number = {6},
pages = {065901},
author = {A Pierret and D Mele and H Graef and J Palomo and T Taniguchi and K Watanabe and Y Li and B Toury and C Journet and P Steyer and V Garnier and A Loiseau and J-M Berroir and E Bocquillon and G Fève and C Voisin and E Baudin and M Rosticher and B Plaçais},
title = {Dielectric permittivity, conductivity and breakdown field of hexagonal boron nitride},
journal = {Materials Research Express},
}

@article{laturiahbn,
  title={Dielectric properties of hexagonal boron nitride and transition metal dichalcogenides: from monolayer to bulk},
  author={Laturia, Akash and Van de Put, Maarten L and Vandenberghe, William G},
  journal={npj 2D Materials and Applications},
  volume={2},
  number={1},
  pages={6},
  year={2018},
  publisher={Nature Publishing Group UK London}
}

@article{kimhbn,
  title={Synthesis and characterization of hexagonal boron nitride film as a dielectric layer for graphene devices},
  author={Kim, Ki Kang and Hsu, Allen and Jia, Xiaoting and Kim, Soo Min and Shi, Yumeng and Dresselhaus, Mildred and Palacios, Tomas and Kong, Jing},
  journal={ACS nano},
  volume={6},
  number={10},
  pages={8583--8590},
  year={2012},
  publisher={ACS Publications}
}

@article{einenkel2011graphene,
  title={Possibility of superconductivity due to electron-phonon interaction in graphene},
  author={Einenkel, Matthias and Efetov, Konstantin B},
  journal={Physical Review B—Condensed Matter and Materials Physics},
  volume={84},
  number={21},
  pages={214508},
  year={2011},
  publisher={APS}
}

@article{eisenstein1992negative,
  title={Negative compressibility of interacting two-dimensional electron and quasiparticle gases},
  author={Eisenstein, JP and Pfeiffer, LN and West, KW},
  journal={Physical review letters},
  volume={68},
  number={5},
  pages={674},
  year={1992},
  publisher={APS}
}

@article{yang2021condensation,
  title={Condensation of collective polar vortex modes},
  author={Yang, Tiannan and Dai, Cheng and Li, Qian and Wen, Haidan and Chen, Long-Qing},
  journal={Physical Review B},
  volume={103},
  number={22},
  pages={L220303},
  year={2021},
  publisher={APS}
}

@book{giuliani2008quantum,
  title={Quantum theory of the electron liquid},
  author={Giuliani, Gabriele and Vignale, Giovanni},
  year={2008},
  publisher={Cambridge university press}
}

@article{pavlenko2021temperature,
  title={Temperature dependence of dielectric properties of ferroelectric heterostructures with domain-provided negative capacitance},
  author={Pavlenko, Maksim A and Tikhonov, Yuri A and Razumnaya, Anna G and Vinokur, Valerii M and Lukyanchuk, Igor A},
  journal={Nanomaterials},
  volume={12},
  number={1},
  pages={75},
  year={2021},
  publisher={MDPI}
}

@article{catalan2012domain,
  title={Domain wall nanoelectronics},
  author={Catalan, Gustau and Seidel, J and Ramesh, Ramamoorthy and Scott, James F},
  journal={Reviews of Modern Physics},
  volume={84},
  number={1},
  pages={119--156},
  year={2012},
  publisher={APS}
}

@article{dassarma2011review,
  title={Electronic transport in two-dimensional graphene},
  author={Das Sarma, Sh and Adam, Shaffique and Hwang, EH and Rossi, Enrico},
  journal={Reviews of modern physics},
  volume={83},
  number={2},
  pages={407--470},
  year={2011},
  publisher={APS}
}

@article{veld2023screening,
  title={Screening induced crossover between phonon-and plasmon-mediated pairing in layered superconductors},
  author={In’T Veld, Y and Katsnelson, MI and Millis, AJ and R{\"o}sner, M},
  journal={2D Materials},
  volume={10},
  number={4},
  pages={045031},
  year={2023},
  publisher={IOP Publishing}
}

@article{Eliashberg1960,
  author = {Eliashberg, G. M.},
  title = {Interactions between electrons and lattice vibrations in a superconductor},
  journal = {Soviet Physics JETP},
  volume = {11},
  pages = {696--702},
  year = {1960}
}

@article{AndoFowlerStern1982,
  author = {Ando, T. and Fowler, A. B. and Stern, F.},
  title = {Electronic properties of two-dimensional systems},
  journal = {Reviews of Modern Physics},
  volume = {54},
  pages = {437--672},
  year = {1982},
  doi = {10.1103/RevModPhys.54.437}
}

@article{randeria2021limits,
  title={Limits on superconductivity in flatland},
  author={Randeria, Mohit},
  journal={Science},
  volume={372},
  number={6538},
  pages={132--132},
  year={2021},
  publisher={American Association for the Advancement of Science}
}
\end{document}